\documentclass[epj]{svjour}
\usepackage[pdftex,colorlinks,citecolor=blue,bookmarks]{hyperref}
\usepackage{graphics}
\usepackage{cite}
\usepackage{color}	
\usepackage{hyphenat}
\usepackage{amsbsy}
\begin{document}

\title{A symmetry conserving description of odd-nuclei with the Gogny force}
\subtitle{Particle number and angular momentum projection with self-consistent blocking}
\author{ M. Borrajo \and J. L. Egido}
%
%
\institute{Departamento de F\'isica Te\'orica, Universidad Aut\'onoma de Madrid, E-28049 Madrid, Spain}
\date{Received: date / Revised version: date}
%
\abstract{
We present an approach for the calculation of odd-nuclei with exact self-consistent blocking and particle number and angular momentum projection with the finite range density dependent Gogny force. As an application we calculate the nucleus $^{31}$Mg at the border of the $N=20$ inversion island.  We evaluate  the ground state properties,  the excited states  and the transition probabilities. In general we obtain a good description of the measured observables.
     } 
\maketitle
\section{Introduction}

\label{intro}

The mean field (MF) and the beyond mean field (BMF) theories \cite{BHR.03,NVR.11} as well as the interacting shell model~\cite{Review_SM} are widely used to describe nuclear structure phenomena.  Traditionally the MF approach has been
chiefly applied to describe global properties of atomic nuclei. Only recently  with the modern BMF methods  the study  of  nuclear spectroscopy started .

 The basic mean field approach, the Hartree\hyp Fock\hyp Bogoliubov (HFB) approach~\cite{RS.80}, allows to deal with single particle as well as  with collective phenomena like the rotations or the superfluidity by the spontaneous symmetry breaking mechanism.  BMF approaches have mainly been developed either in small configuration spaces and using shell-model-like interactions \cite{SG.87,Sch.04} or in large configuration spaces and employing density-dependent interactions \cite{BHR.03}.  The  BMF ingredients are two. On the one hand the recovery of the symmetries broken in the HFB approach, like particle number (PN) and angular momentum (AM) projection, and on the other hand the incorporation of  fluctuations around the most probable MF values in the frame of the generator coordinate method (GCM). The combination of these two methods in a unified framework is the so-called symmetry conserving configuration mixing (SCCM) method.
 The best current SCCM calculations~\cite{PRC_78_024309_2008,PRC_81_044311_2010,PRC_81_064323_2010} include the quadrupole (axial and triaxial) deformations as generator coordinates. The recovery of the PN symmetry is carried out~\cite{RS.80} either in the projection after variation (PAV) approach~\cite{PRC_78_024309_2008,PRC_81_044311_2010,PRC_81_064323_2010} or in the variation after projection (VAP) approach~\cite{PRC_81_064323_2010}. The AM projection is always performed in the PAV approach.  An awkward feature of the AM-PAV is a stretching of the whole spectrum \cite{PRL_99_062501_2007}. To correct this problem the cranking frequency has been recently incorporated  as a coordinate \cite{BRE.15,EBR.16} showing the relevance of this degree of freedom to provide a quantitative description of spectra and transition probabilities.  These calculations, however, require long CPU time in modest computer facilities.  A more CPU friendly approach is provided by the Bohr Hamiltonian approach  \cite{RS.80}, broadly used recently \cite{Liebert_Gogny,pole_PQQ,Ring_Rel,Japs}.

 These developments have taken place for even-even nuclei and it seems natural to extend this type of approaches to odd-even and odd-odd nuclei. As a matter of fact angular-momentum projected calculations  for odd-A nuclei started long ago, though they have been mostly performed on HF or HFB states  in small valence spaces \cite{BGR.65,GW.67,RPK.93,HI.84,HSG.85}. More recently a GCM mixing based on parity and AM projected  Slater determinants in a model space of antisymmetrized Gaussian wave packets has been carried out in the frameworks of antisymmetrized \cite{KTK.13,KK.10} and fermionic \cite{NF.08} molecular dynamics.  Very recently  in  Ref.~\cite{Bally}   calculations for the nucleus $^{25}$Mg have been performed  using exact blocking in the  $(\beta, \gamma)$-constrained  HFB wave functions which were used to generate the GCM states. Subsequently these states were projected to angular momentum and particle number to provide the definitive variational space to solve the Hill-Wheeler-Griffin equation. In this calculation the Skyrme parameterization  SLyMR0  was used.  

The Gogny force has proven along the years to be a very reliable choice with predictive power and as a matter of fact is being used as a benchmark for other calculations. It seems therefore desirable to extend the BMF approaches used for even-even to odd systems with this interaction \cite{PRC_81_064323_2010,EBR.16}. In Ref.~\cite{BE.16} we proposed the  use of the finite range density dependent Gogny force \cite{DG.80} to calculate ground state properties of odd nuclei in a symmetry conserving approach.

In this paper we report on  calculations with particle number and angular momentum projection for the description of the nucleus $^{31}$Mg with the Gogny force.   Our starting point is the exact self-consistent blocked HFB theory. The recovery of the particle number symmetry (and the associated handling of the pairing correlations) that we consider as  basic ingredient is performed in the variation after projection (VAP) approach.  The nonlinearity of the projected HFB equations hinders, due to the huge CPU time needs,  to carry out an angular momentum projection in a VAP  approach.  However, in order to include as much as possible angular momentum effects in the variation we perform a restricted AM-VAP by considering the deformation parameters $(\beta,\gamma)$  as coordinates to generate a collective space, where, after angular momentum projection,  the minimum is determined.  The nucleus $^{31}$Mg chosen as example of our approach  is specially interesting  because it is a landmark of the inversion island in the $N=20$ region. 
Therefore it is a big challenge for any new approach to describe correctly the properties of this nucleus. We would  like also to mention the limitations of the present approach: In a symmetry conserving mean field approach one does not consider fluctuations around the most probable values, i.e., we are not considering the large amplitude fluctuations of the GCM. Furthermore in our description we only consider one-quasiparticle states.   The nucleus $^{31}$Mg has been  thoroughly discussed theoretically \cite{Ki.07,CNP.14} and experimentally \cite{Ne.11}.

In Sect.~\ref{Sect:Theo} we describe the basics of our theoretical method.  The first results of the calculations are shown in Sect.~\ref{Sect:PES}, the potential energy surfaces (PES) for different angular momenta are displayed in Subsect.~\ref{Sect:PES_PNAMP}. The spectrum of $^{31}$Mg  is discussed in Sect.~\ref{Sect:spec_plus} together with the electromagnetic properties of this nucleus. Finally, the conclusions and outlook are presented 
in Sect.~\ref{Sect:concl}.

\section{Theoretical approach}
\label{Sect:Theo}
Our starting point is the Bogoliubov transformation 
  \begin{equation}  \label{bogtrans}
  \alpha _\mu =\sum_kU_{k\mu }^{*}c_k+V_{k\mu }^{*}c_k^{\dagger},
  \end{equation}
  with ${c_{k}^{\dagger},c_{k}}$ the particle creation and annihilation operators in the original basis, for example in the triaxial harmonic oscillator one. $U$ and $V$ are the Bogoliubov matrices  to be determined by the Ritz variational principle.

In the HFB theory \cite{Ma.75,RS.80}  the wave function of the ground state of an even-even  nucleus is a product wave function of the form
\begin{equation}
|\phi \rangle_{e-e} = \prod_{\mu} \alpha_{\mu} |-\rangle,
\end{equation}  
where the states of positive and negative simplex, see below,  appear pairwise \cite{EMR.80}.
Analogously the wave function  for an even-odd nucleus is given by
\begin{equation}
|\phi \rangle_{o-e} = \alpha^{\dagger}_b |\phi \rangle_{e-e},
\label{wf_e_o}
\end{equation}  
where the  symbol $b$ characterizes the quantum numbers of the blocked level.

  Since the Bogoliubov transformation mixes creator and annihilator operators the HFB wave functions  ($|\phi \rangle_{e-e}$ or $|\phi \rangle_{o-e}$), in general,  are not eigenstates of the particle number operator. 
   Furthermore if  the  index $k$, in Eq.(\ref{bogtrans}), is allowed to run indiscriminately over all
   states of the basis, all symmetries of the system such as parity, angular momentum etc.  are broken.
   Since in this work we will deal only with odd-even nuclei we suppress the subscript ${o-e}$ in the following.

  To recover the broken symmetries  the projection technique is very convenient, in particular the wave function
 \begin{equation}
 |\Phi_{M}^{N,I}\rangle =  \sum_K g^{I}_K \hat{P}^I_{MK} P^N P^Z| \phi \rangle,
 \label{Proj_WF}
 \end{equation}
 is an eigenstate of the particle number and the angular momentum operators.  Where we have introduced the   projectors on the particle number (PNP), $P^N$, and the angular momentum (AMP), $P^I_{MK}$, respectively.    
   In the following equations $P^N$ stays for $P^NP^Z$.
 The $g_K$ parameters have to be determined by the variational principle \cite{RS.80}, see below.  
 The operator $\hat{P}_{MK}^{I}$ is the angular momentum
 projection operator \cite{RS.80} given by
 
 \begin{eqnarray}
 \hat{P}^{I}_{MK}&=&\frac{2I+1}{8\pi^{2}}\int\limits_{0}^{2\pi}d\boldsymbol{\gamma}\int\limits_{0}^{\pi}d
 \boldsymbol{\beta}\sin(\boldsymbol{\beta)} \nonumber \\
 & & \times \int\limits_{0}^{2\pi}d\boldsymbol{\alpha} D^{I*}_{MK}(\boldsymbol{\alpha},\boldsymbol{\beta},\boldsymbol{\gamma})\hat{R}(\boldsymbol{\alpha},\boldsymbol{\beta},\boldsymbol{\gamma}), 
\label{AMProj}
 \end{eqnarray}
  where $\hat{R}(\boldsymbol{\alpha},\boldsymbol{\beta},\boldsymbol{\gamma})=e^{-i\boldsymbol{\alpha}\hat{J}_{z}}e^{-i\boldsymbol{\beta}\hat{J}_{y}}e^{-i\boldsymbol{\gamma}\hat{J}_{z}}$ is the rotation operator,  
 $D^{I}_{MK}(\boldsymbol{\alpha},\boldsymbol{\beta},\boldsymbol{\gamma})$ is the Wigner function and   ($\boldsymbol{\alpha}$,$\boldsymbol{\beta}$,$\boldsymbol{\gamma}$) 
 are the Euler angles. We use bold Greek letters for the Euler angles ($\boldsymbol{\alpha},\boldsymbol{\beta}$,$\boldsymbol{\gamma}$)  and non-bold for the quadrupole deformation parameters $(\beta,\gamma$). We use 32 integration points for each Euler angle
 in the full integration interval $\boldsymbol{\alpha}\in [0,2\pi]$, $\boldsymbol{\beta}\in [0,\pi]$, $\boldsymbol{\gamma}\in [0,2\pi]$. $I, M, K$ are the total angular momentum and its projection on the $z$ axis in the laboratory and intrinsic frame, respectively.

 The particle number operator is given by
 \begin{equation}{\hat{P}}^N  =  \frac{1}{2\pi}\int_{0}^{2\pi} 
 {e}^{i \varphi (\hat{N}-N)} \, d{\varphi},
 \label{eq:PN}
 \end{equation}
 the variable $\varphi$ is the canonical conjugated coordinate to $\hat{N}$ in the associated gauge space.
The PN-projection involves a rotation over the gauge angle which is discretized \cite{Fo.70} using 11 points in the interval $[0,2\pi]$ .  

To fix  the symmetry conserving wave function of Eq.~(\ref{Proj_WF}) one has to determine the matrices $\{U,V\}$
and the coefficients $g$. There are several ways to attain this.  The optimal way is to minimize the PN and AM projected 
energy with respect to $\{U,V\,g\}$.  This is the variation after projection method (VAP), in short PNAM-VAP.   This method as mentioned in the Introduction has the drawback of its 
high CPU time consumption. Another possibility is to minimize the non-projected, i.e., the HFB, energy and to perform afterwards the PN and AM projection, i.e., the so-called projection after variation (PAV) method, or  PNAM-PAV in short. This approach has the  disadvantage that since the superfluid phase is not very collective  a pairing collapse can take place.  A CPU-affordable solution would be to perform a PN-VAP approach and afterwards an AM-PAV one.  This approach, however, has the disadvantage that the minimum determined in this way is insensitive to the angular momentum. The solution we adopt in this research  is to perform a PN-VAP approach but exploring the relevant degrees of freedom of the system.  In other words we perform an approximated  AM-VAP in the sense that we search for an AM-VAP minimum in a restricted collective configuration space for each angular momentum. In this work we consider the deformation parameters $(\beta,\gamma)$ as the additional degrees of freedom which span the collective variational space. 

To create wave functions corresponding to different values of the parameters $(\beta,\gamma)$
we minimize the constrained PN-VAP energy
\begin{eqnarray}
{E^{\prime}}[\phi]= \frac{ \langle\phi^{}|\hat{H}\hat{P}^{N}|\phi{} \rangle}{\langle\phi^{}|\hat{P}^{N}|\phi^{} \rangle} -  \langle \phi |\lambda_{q_{0}}\hat{Q}_{20} + \lambda_{q_{2}}  \hat{Q}_{22} | \phi \rangle, \label{E_Lagr_bet-gam}
\end{eqnarray}
with  the Lagrange multipliers $\lambda_{q_{0}}$  and $\lambda_{q_{2}}$ being determined by the constraints 
\begin{equation}
\langle \phi |\hat{Q}_{20} | \phi \rangle =q_{0}, \;\; \; \langle \phi |\hat{Q}_{22} | \phi \rangle =q_{2}. \label{q0_q2_constr}
\end{equation}
The relation between $(\beta,\gamma)$ and $(q_{0},q_{2})$ is provided by Eqs.~(\ref{be_ga0},\ref{be_ga}).
\begin{eqnarray}
\beta &=&\frac{1}{3r^{2}_{0}A^{5/3}} \sqrt{20\pi(\langle\hat{Q}_{20}\rangle^{2} + 2 \langle \hat{Q}_{22}\rangle^{2})} \label{be_ga0},\\
\gamma &=& \arctan\left(\sqrt{2}\frac{\langle \hat{Q}_{22}\rangle}{\langle\hat{Q}_{20}\rangle}\right), \label{be_ga}
\end{eqnarray}
with $r_{0}=1.2$ fm and  $A$ the mass number. The minimization of Eqs.~(\ref{E_Lagr_bet-gam}-\ref{q0_q2_constr}) is performed with the conjugated gradient method \cite{grad}. The blocking structure of the wave function of Eq.~(\ref{wf_e_o}) is a self-consistent symmetry and for a given blocking number we determine the lowest solution in the blocked channel compatible with the imposed constraints. In our calculations we impose three discrete self-consistent symmetries, namely, parity ($\hat{P}$), simplex  ($\Pi_1=\hat{P}e^{-i\pi J_x}$) and the 
 $\Pi_2\mathcal{T}$ symmetry with  $\Pi_2=\hat{P}e^{-i\pi J_y}$ and $\mathcal{T}$ the time reversal operator. The first two symmetries provide good parity and simplex quantum numbers and the third allows to use only real quantities.  Therefore we have in principle eight blocking channels, protons with positive, $\pi^{+}$,  or negative  parity, $\pi^{-}$, and additionally each of them ($\pm i$) simplex and the corresponding  $(\nu^{+}, \nu^{-})$ for neutrons. However, since we are not doing cranking calculations the states with positive or negative simplex are practically degenerate and we only block one of them. We have therefore four blocking channels $(\nu^{+}, \nu^{-}, \pi^{+}, \pi^{-})$, furthermore for an odd neutron nucleus like $^{31}$Mg they reduce to two, namely  $\nu^{+}$ and $\nu^{-}$. Once we have chosen the blocking channel to solve Eqs.~(\ref{E_Lagr_bet-gam},\ref{q0_q2_constr}) we iterate until the lowest energy is found.

The next step is the simultaneous particle number and angular momentum projection of each state   $|\phi(\beta,\gamma)\rangle$  that conforms the grid, 
\begin{eqnarray}
|\Psi^{N,I}_{M,\sigma } (\beta,\gamma) \rangle &=&   \sum_{K} g^{I}_{K\sigma} \hat{P}^N \hat{P}^I_{MK} \; |\phi (\beta,\gamma)\rangle  \nonumber \\
& =&   \sum_{K} g^{I}_{K\sigma}  |IMK,N, (\beta,\gamma) \rangle,  \label{eq:GCM_Ansatz_bet_gam}
\end{eqnarray}
where the  coefficients  $g^{I}_{K\sigma}$  are variational parameters.  They are determined by the energy minimization which  provides a reduced Hill-Wheeler-Griffin equation 
\begin{equation}
\sum_{K'} \, \,(\mathcal{H}^{N,I}_{K, K'} - E^{N,I}_\sigma \mathcal{N}^{N,I}_{K,K'}) g^{I}_{K^{\prime}\sigma} = 0,
\label{HW_K}
\end{equation} 
where $\mathcal{H}^{N,I}_{K K'}$ and $\mathcal{N}^{N,I}_{K,K'}$ 
are the Hamiltonian and norm overlaps defined by 
\begin{eqnarray}
\mathcal{H}^{N,I}_{K, K'} \! & = & \!   \langle IMK,N, (\beta,\gamma) |\hat{H} | IMK',N,(\beta,\gamma) \rangle \label{hamove} \\
\mathcal{N}^{N,I}_{K,K'} \! & = & \!   \langle IMK,N,(\beta,\gamma) | IMK',N,(\beta,\gamma) \rangle \label{normove}.
\end{eqnarray}
The presence of the norm matrix in Eq.~(\ref{HW_K}) is due to the non-orthogonality of the  states $|IMK,N,(\beta,\gamma) \rangle$. Eq.~(\ref{HW_K}) is solved by standard techniques \cite{RS.80}.
Notice that in each $(\beta,\gamma)$ point one can have several eigenvalues $E^{N,I}_\sigma$ labeled by $\sigma$. 

Since the states $|IMK,N,(\beta,\gamma) \rangle$ are not orthogonal the weights  $g^{I}_{K\sigma}$
do not satisfy  $\sum_{K} |g^{I}_{K\sigma}|^2=1$. The collective wave function
\begin{equation}
G^{I}_{K,\sigma} = \sum_{K'} \large( \mathcal{N}^{N,I}\large)_{K,K'}^{1/2} g^{I}_{K',\sigma}, 
\label{G_KK}
\end{equation}
on the other hand, does and can be interpreted as a probability amplitude.

To clarify our procedure  we sketch the different steps of the calculations. Our goal is to find the energy minimum for a given spin and parity using the wave function of  Eq.~(\ref{eq:GCM_Ansatz_bet_gam}). That means, we have to find the $(\beta,\gamma)$ values $(\beta^{I}_{\rm min},\gamma^{I}_{\rm min})$ such that $|\Psi^{N,I}_{M,\sigma } (\beta^{I}_{\rm min},\gamma^{I}_{\rm min}) \rangle$ provides the 
energy minimum for the given spin and parity. 

{\em Step 1:} We choose the desired parity for the blocked state in Eq.~(\ref{wf_e_o}). Next we solve the PN-VAP variational equations Eqs.~(\ref{E_Lagr_bet-gam}, \ref{q0_q2_constr}) for all 
	$(\beta,\gamma)$ values of the grid. This step provides a set of wave functions $\phi(\beta,\gamma)$ with the right  parity but without any angular momentum content.

{\em Step 2:}  To find out $(\beta^{I}_{\rm min},\gamma^{I}_{\rm min})$ we now solve Eq.~(\ref{HW_K}) for all  $\phi(\beta,\gamma)$ of the grid determined in Step 1 for a given $I$-value.  The minimum  value of $E^{N,I}_{\sigma=1}(\beta,\gamma)$ provides the  $(\beta^{I}_{\rm min},\gamma^{I}_{\rm min})$ values. 

{\em Step 3:}  Repeat Step 2 for all $I$-values.  The energies  $E^{N,I}_{\sigma=1}(\beta^{I}_{\rm min},\gamma^{I}_{\rm min})$ allow to draw the spectrum and the wave functions  $|\Psi^{N,I}_{M,\sigma } (\beta^{I}_{\rm min},\gamma^{I}_{\rm min}) \rangle$ enable the calculation of electromagnetic properties or any other observable.

In the calculations we use the Gogny interaction \cite{DG.80} with the D1S parameterization \cite{BGG.91}.
In order to avoid potential problems with the PNP  we consider all exchange terms of the force interaction, the Coulomb force and the two-body correction of the kinetic energy  \cite{AER.01E}. Concerning the density dependence of the force we adopt  the projected density prescription for the PNP which has been proven to provide divergence free results \cite{AER.01P}.  For the AMP we use  the mixed density prescription which provide perfectly convergent calculations \cite{Nuria2}.  For further details see for example Ref.\cite{PRC_81_064323_2010}. A very detailed discussion of  these aspects has  recently been presented in Ref.~\cite{EG.16}.

\section{Results: The potential energy surfaces}
\label{Sect:PES}
As a numerical application we apply this theory to the calculation of several properties of the nucleus $^{31}$Mg. In the calculations the intrinsic many body wave functions $|\phi(\beta,\gamma)\rangle$ are expanded in a Cartesian harmonic oscillator basis and the number of spherical shells included in this basis is $N_{shells}=8$ with an oscillator length of $b=1.01A^{1/6}$. We have compared with calculations with  10 shells and we find a good convergence.

\begin{figure}[t]
	\begin{center}
		\resizebox{0.48\textwidth}{!}{%
			\includegraphics{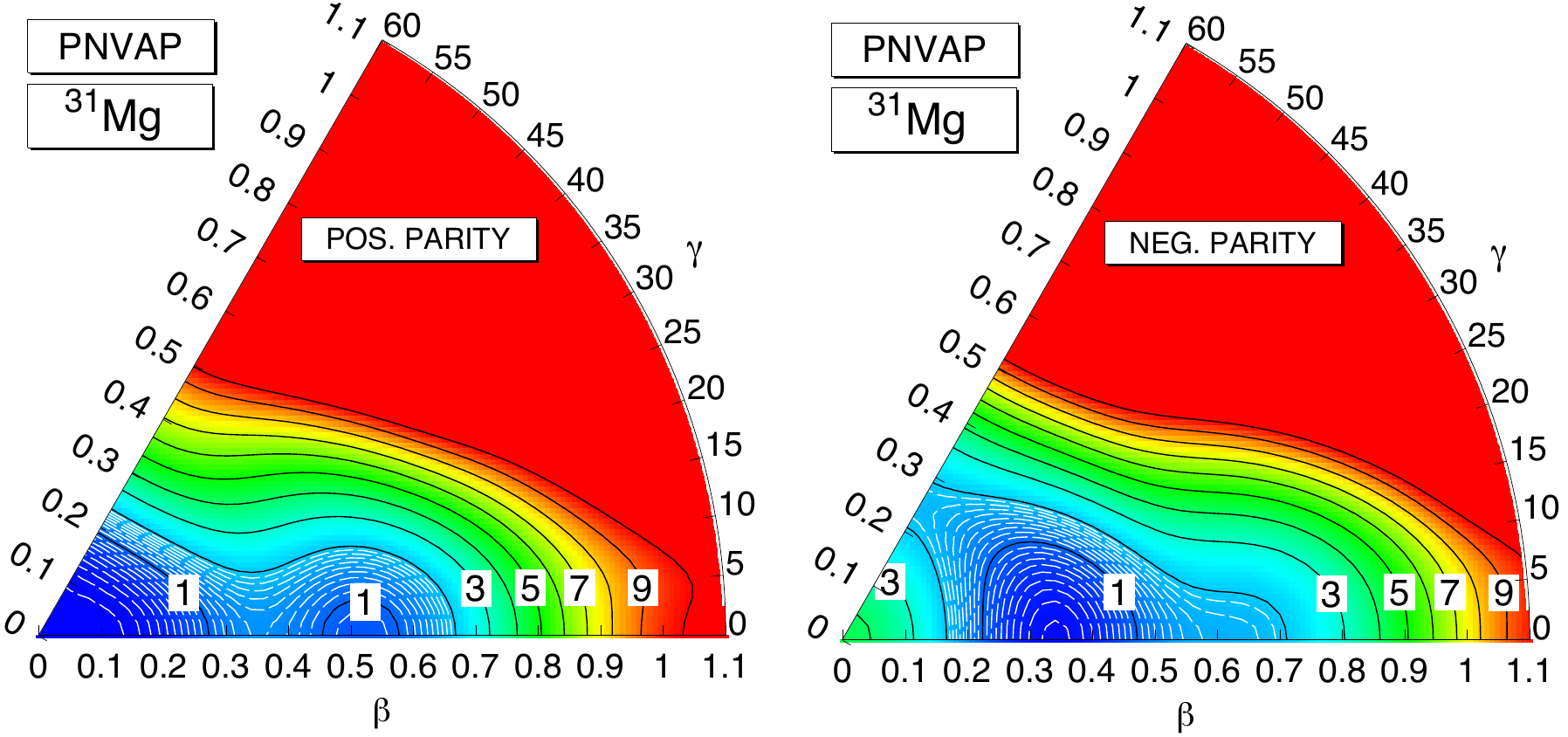}}
		\caption{Potential energy surfaces (in MeV) of $^{31}$Mg  in the PNAMP method for positive (left) and negative (right)  parity states.The energy origin has been chosen independently for each panel and the energy minimum has been set to zero.   The continuous lines represent contours from 1 to 10 MeV in 1 MeV steps.   The white dahed contours around the minima are 0.1 MeV apart and extend from 0.1 up to 1.9 MeV. The absolute value of the energy is $-243.472$ MeV (positive parity) and $-243.129$ MeV (negative parity)}
		\label{Fig:PES_VAP}       
	\end{center}
\end{figure}

\begin{figure*}[t]
	\begin{center}
		\resizebox{1.0\textwidth}{!}{%
			\includegraphics{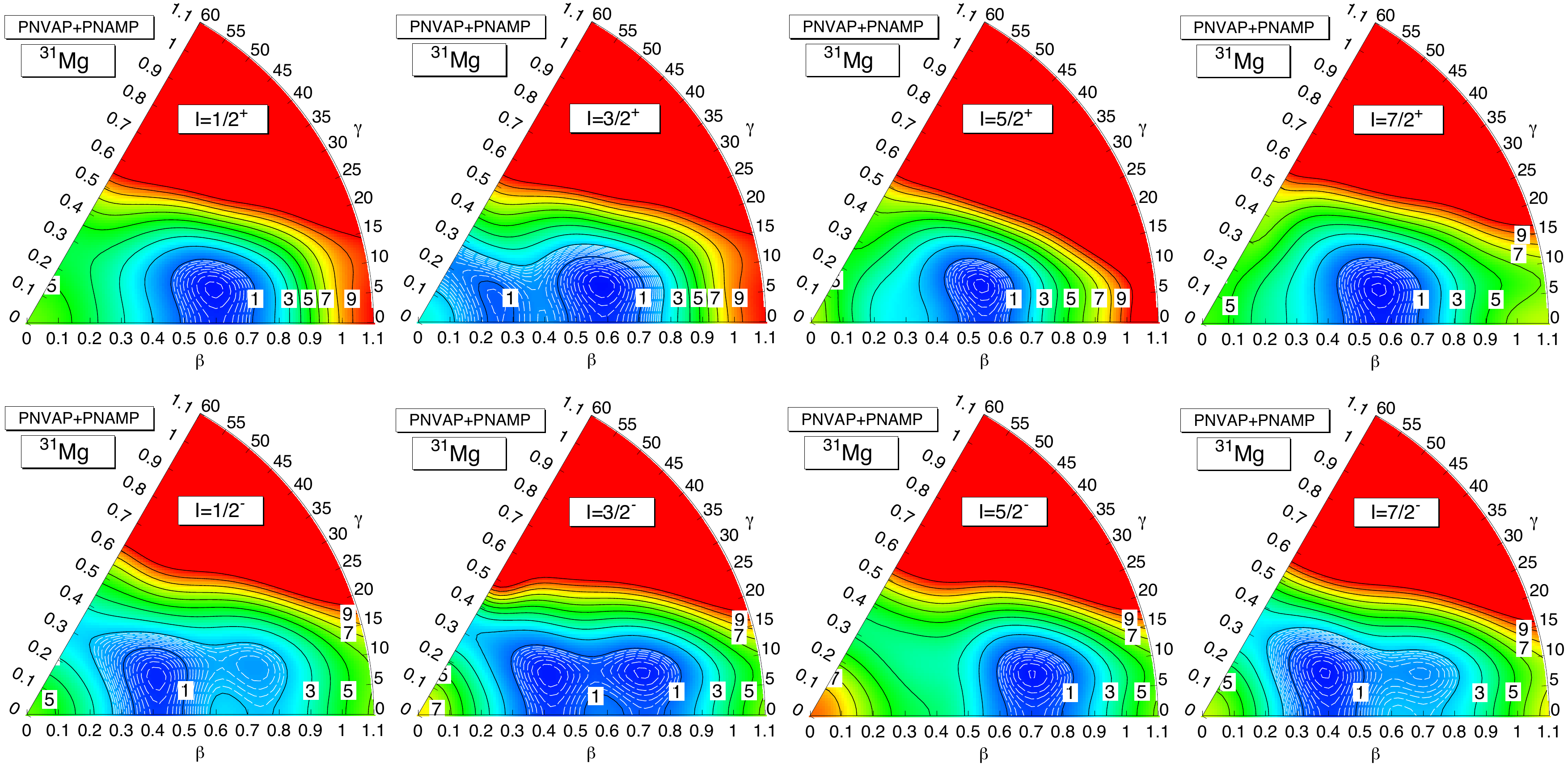}}
		\caption{Potential energy surfaces (in MeV) of $^{31}$Mg  in the PNAMP method for positive (top panels) and negative (bottom panels)  parity states.The energy origin has been chosen independently for each panel and the energy minimum has been set to zero.  The continuous lines represent contours from 1 to 10 MeV in 1 MeV steps.   The white dashed contours around the minima are 0.1 MeV apart. The absolute values (in MeV) of the energy minima are $(-246.840, -246.553, -245.791, -245.0127)$ for angular momenta and parity $\frac{1}{2}^{+}, \frac{3}{2}^{+},\frac{5}{2}^{+},\frac{7}{2}^{+}$ and  $(-244.282, -246.152, -245.557, -245.990)$ for  $\frac{1}{2}^{-}, \frac{3}{2}^{-},\frac{5}{2}^{-},\frac{7}{2}^{-}$  respectively.}
		\label{Fig:PES_PNAMP}       
	\end{center}
\end{figure*}

\subsection{The PN-VAP potential energy surfaces}
\label{Sect:PES_VAP}

The solution of  Eq.~(\ref{E_Lagr_bet-gam})  in the $(\beta,\gamma)$ plane  for 99 points in a grid of triangles provides PN-VAP wave functions as a function of the deformation parameters. The quantities
  \begin{eqnarray}
 {E}(\beta,\gamma)= \frac{ \langle\phi^{}(\beta,\gamma)|\hat{H}\hat{P}^{N}|\phi{}(\beta,\gamma) \rangle}{\langle\phi^{}(\beta,\gamma)|\hat{P}^{N}|\phi^{} (\beta,\gamma)\rangle} \label{Eq:PES_VAP} \end{eqnarray}
 as a function of $(\beta,\gamma)$ provide potential energy surfaces.  These are plotted in
 Fig.~\ref{Fig:PES_VAP} for the nucleus $^{31}$Mg with 12 protons and 19 neutrons for 
 blocked positive (negative) parity neutron states in the left (right) hand panel.  In the positive parity
 channel we find two minima on the prolate axis, about 300 keV apart, one at $\beta\approx 0.08$ and the other at
 $\beta\approx 0.5$.  Along the prolate axis the surface is softer than along the oblate one. For $\beta$ values
 smaller than 0.3 the PES is rather soft in the $\gamma$ degree of freedom. In the negative parity channel 
 we observe one clear minimum at $\beta\approx 0.34$ on the prolate axis and an incipient secondary minimum at $ \beta\approx 0.6$. We also observe a saddle point at $\beta\approx 0.25$ on the oblate axis. The softening of PES in the $\gamma$ degree of freedom now extends to larger values than for the positive parity.  Comparing both parities one observes  that  the minima of the  negative parity channel are shifted to larger deformations as compared to the positive one.

 \subsection{The PNAMP  potential energy surfaces}
 \label{Sect:PES_PNAMP}
 
The  solutions of  Eq.~(\ref{E_Lagr_bet-gam}) for the different $(\beta,\gamma)$ values have been determined without constraint on the angular momentum.  For odd nuclei, at variance with even-even nuclei where all solutions for different  $(\beta,\gamma)$ values satisfy $\langle J_x \rangle=0$, each point of the PES 
can have different $\langle J_x \rangle$. 

 In order to obtain meaningful PES one has to solve the Hill-Wheeler-Griffin equations, Eq.~(\ref{HW_K}), for different values of the angular momentum.    In Fig.~\ref{Fig:PES_PNAMP} we display the results of such calculations.
  As mentioned above at each $(\beta,\gamma)$ point the Hill-Wheeler-Griffin equation provides several solutions, which we have numbered by the symbol $\sigma$. In particular for a given $I$ there are $(2I+1)/2$ linearly independent solutions. The PES displayed in Fig. ~\ref{Fig:PES_PNAMP} have been made taking at each point the lowest solution of Eq.~(\ref{HW_K}).
  In the top (bottom) panels the results for positive (negative) parity states are shown. In the top panels we present the PESs for angular momenta $\frac{1}{2}^+,\frac{3}{2}^+,\frac{5}{2}^+,\frac{7}{2}^+$, these values being used to label  the corresponding panels.    With the exception of the PES for  $I=\frac{3}{2}^+$ which has two coexisting minima, the other PESs do have only one minimum.  This is somewhat in contrast to the positive parity PES of the PN-VAP approach of Fig.~\ref{Fig:PES_VAP} which has two. All four PESs present a clear triaxial minimum at $\beta =0.61$ and $\gamma=14^{\circ}$.  This common minimum is well localized and  much softer in the $\beta$ than in the $\gamma$ degree of freedom.  The secondary minimum for $I=\frac{3}{2}^+$  is localized at $\beta=0.31$ and $\gamma=14^{\circ}$ and appears at slightly higher energy.  This secondary  minimum in contrast to the primary one is very soft in the $\gamma$ degree of freedom.  The exact numerical values of the deformations and the excitation energies are given in Table~\ref{table1}.
 
\begin{figure}[t]
	\begin{center}
		\resizebox{0.45\textwidth}{!}{%
			\includegraphics{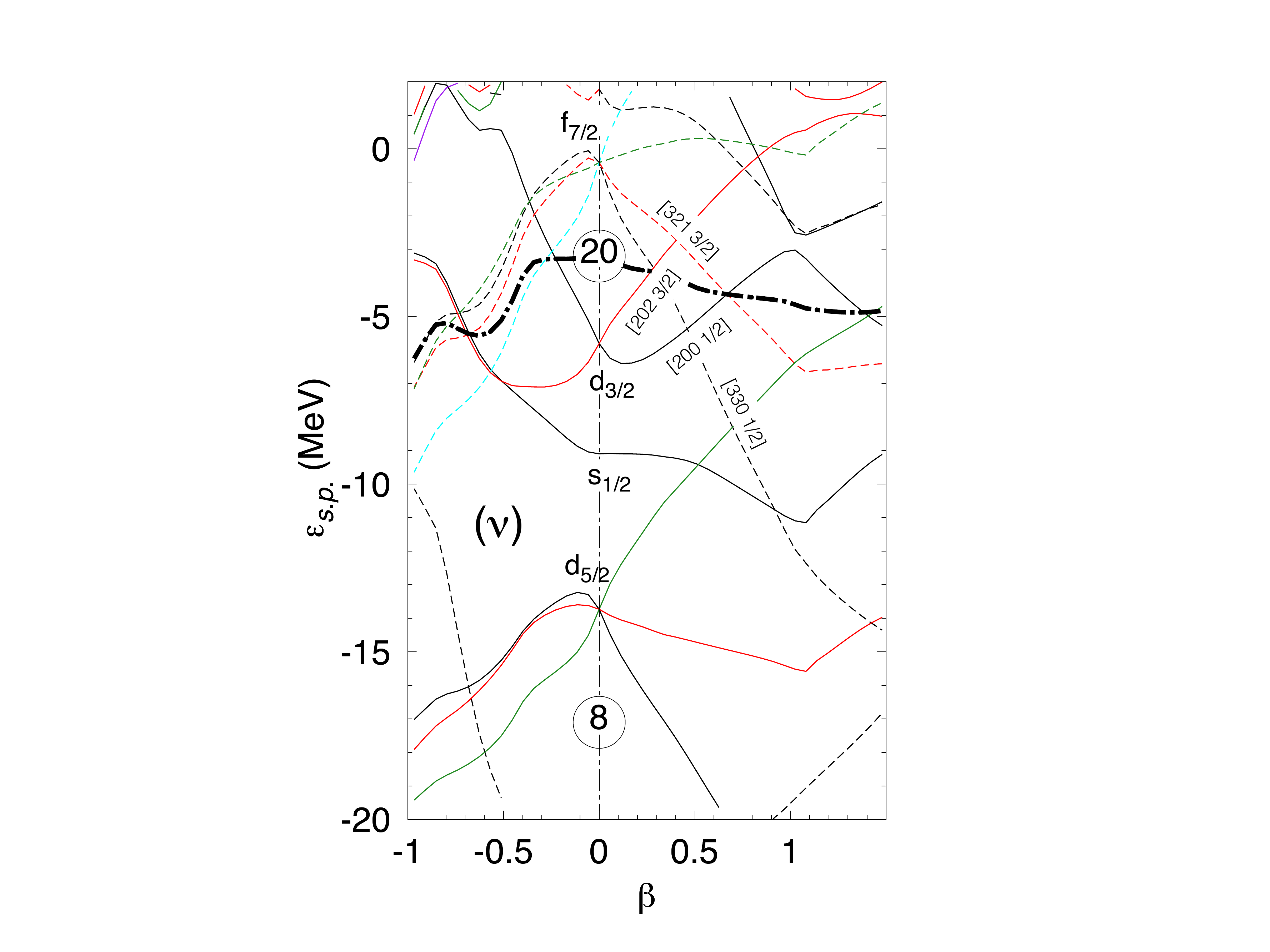}}
		\caption{Single-particle levels of $^{32}$Mg for neutrons in the HFB approach. The thick dashed line represents the Fermi level. The Nilsson quantum numbers $[N,n_{z},m_{l},\Omega]$ are indicated for the relevant orbitals.}
		\label{SPE_neu}       
	\end{center}
\end{figure}

To identify the dominant configuration associated to each minimum it is convenient to consider the single particle levels involved around the Fermi level. In Fig.~\ref{SPE_neu}  we display the neutron single particle energies in the HF approach for $^{32}$Mg, which we will use to provide qualitative arguments.  In  $^{31}$Mg and for positive parity states we must have one particle either in the Nilsson orbital  $[200\frac{1}{2}]$ or  $[202\frac{3}{2}]$.  Furthermore, one can get more insight looking at the  $K$-distribution (the $G^{I}_{K\sigma}$ coefficients), see Eq.~(\ref{G_KK}), of the wave function in the energy minimum, shown in Table~\ref{table1}. As we can see the $I =\frac{3}{2}$ state with $\beta=0.31$, $\gamma=14^{\circ}$ corresponds to  $|K| =\frac{3}{2}$ and the $I =\frac{3}{2}$ state with $\beta=0.61$, $\gamma=14^{\circ}$ to 
 $|K| =\frac{1}{2}$.  That means in the minimum with smaller deformation the blocked particle sits in the Nilsson orbital with  $[202\frac{3}{2}]$ whereas  in the minimum with the large deformation the blocked particle sits in the $[200\frac{1}{2}]$.  Looking at the single particle energies of these levels it is obvious that in the first case the two additional particles are occupying the $[200\frac{1}{2}]$ level.  This is the normal occupation that one expects.  In the second case, when the blocked particle is in the  $[200\frac{1}{2}]$,  the level $[202\frac{3}{2}]$ has crossed the Fermi level, see Fig.~\ref{SPE_neu},  and the two additional particles sit in the f$_\frac{7}{2}$ orbit.
 This is the intruder occupation that one observes in the inversion island. Since in our calculations the ground state  corresponds to this configuration we conclude that in our theory the nucleus $^{31}$Mg is inside the inversion region. The large deformation $(\beta=0.61)$ as well as the fact that  the particle sits in the d$_\frac{3}{2}$ orbital, i.e., small Coriolis interaction, is a clear indication that we are in the strong coupling limit,  and that therefore  we must  have $I= \frac{1}{2}$ \cite{RS.80}.  Independently of the Nilsson plot  of Fig.~\ref{SPE_neu} we  can check the occupation of the orbits in the canonical basis of the PN-VAP solution. We find that in the shell model language the minimum at small $\beta$ value corresponds mostly to  $({\rm d}_{\frac{3}{2}})^3({\rm f}_{\frac{7}{2}})^0$, i.e., a 0p1h configuration. The large $\beta$ value corresponds mostly to  $({\rm d}_{\frac{3}{2}})^1({\rm f}_{\frac{7}{2}})^2$, i.e., a 2p3h configuration.

We now discuss the negative parity channel, i.e., the lower panels of Fig.~\ref{Fig:PES_PNAMP}. We observe minima with two deformations for the different spins, see also Table~\ref{table1}, one at $\beta=0.45,\gamma=19^{\circ}$ (spins $\frac{1}{2}^-$, $\frac{3}{2}^-$, $\frac{7}{2}^-$) and the other at $\beta=0.69,\gamma=12^{\circ}$ (spins $\frac{3}{2}^-$, $\frac{5}{2}^-$). Again as for $I=\frac{3}{2}^+$  we find coexistent shapes. In Table~\ref{table1} we also find that the first one corresponds to $|K|=\frac{1}{2}$ and the second one to  $|K|=\frac{3}{2}$.  For negative parity  the blocked particle must sit in the f$_\frac{7}{2}$ orbit. If we look at Fig.~\ref{SPE_neu} we observe that the relevant levels in this orbit are the $[330\frac{1}{2}]$ and the $[321\frac{3}{2}]$. The first level appears at small deformation and will explain the first minimum and the second level at larger deformation  the second minimum.  In both cases we have two additional particles either in the  d$_\frac{3}{2}$ (normal occupation, $\beta=0.45$ minimum) or in the   f$_\frac{7}{2}$ (intruder occupation, $\beta=0.69$ minimum). In the shell model language the minimum at small $\beta$ value corresponds mostly to  $({\rm d}_{\frac{3}{2}})^2({\rm f}_{\frac{7}{2}})^1$, i.e., a 1p2h configuration. The large $\beta$ value corresponds mostly to  $({\rm d}_{\frac{3}{2}})^0({\rm f}_{\frac{7}{2}})^3$, i.e., a 3p4h configuration.  

  \begin{table}
  	\centering
  	\setlength\tabcolsep{4.5pt}
  	\begin{tabular}{|c|c | c| c| c| c|}
  		\hline 
  		$		\footnotesize I^{\pi}_{\sigma}  $  &  $\beta ,\gamma$ &  $E^{+}_{\sigma}$ &   $\vert K \vert=\frac{1}{2}$ &
  		$\vert K \vert=\frac{3}{2}$ &  $\vert K \vert=\frac{5}{2}$
  		\tabularnewline
  		\hline 
  		${1}/{2}^{+}_{1}$  &  $0.61,13.9^{\circ}$ &   $0$ &  $100$ &  $-$ &  $-$
  		\tabularnewline
  		${3}/{2}^{+}_{1}$  &  $0.61,13.9^{\circ}$ &   $0.287$ & $99.4$ &  $0.6$ &  $-$ 
  		\tabularnewline
  		${3}/{2}^{+}_{2}$  &  $0.31,13.9^{\circ}$ &   $0.848$ & $10.6$ &  $89.4$ &  $-$
  		\tabularnewline
  		${5}/{2}^{+}_{1}$  &  $0.61,13.9^{\circ}$ &   $1.049$ &$97.7$ &  $2.3$ &  $0.0$
  		\tabularnewline
  		${7}/{2}^{+}_{1}$  &  $0.61,13.9^{\circ}$ &   $1.827$ &$98.2$ &  $1.7$ &  $0.1$
  		\tabularnewline
  		\hline 
  		${1}/{2}^{-}_{1}$  &  $0.45,19.1^{\circ}$ &   $2.558$ &$100$ &  $-$ &  $-$
  		\tabularnewline
  		${3}/{2}^{-}_{1}$  &  $0.45,19.1^{\circ}$ &   $0.688$ &$99.6$ &  $0.4$ &  $-$
  		\tabularnewline
  		${3}/{2}^{-}_{2}$  &  $0.69,12.2^{\circ}$ &   $0.745$ & $0.1$ &  $99.9$ &  $-$
  		\tabularnewline
  		${5}/{2}^{-}_{1}$  &  $0.69,12.2^{\circ}$ &   $1.283$ & $0.1$ &  $99.9$ &  $0.0$
  		\tabularnewline
  		${7}/{2}^{-}_{1}$  &  $0.45,19.1^{\circ}$ &   $0.850$ & $97.7$ &  $2.0$ &  $0.3$
  		\tabularnewline
  		\hline  
  		
  	\end{tabular}
  	\caption{Properties of the minima of the PESs of Fig.~\ref{Fig:PES_PNAMP} in the different columns. 1: Spin and parity, 2: $(\beta,\gamma)$ coordinates of the minima, 3: Excitation energy (in MeV) with respect to the $I=\frac{1}{2}^{+}$ state, 4, 5, 6: The weights $|G^{I}_{K\sigma}|^{2}$, in percentage, see Eq.~(\ref{G_KK}), for different $|K|$ values.}
  	\label{table1}
  \end{table}

 \begin{figure}[t]
 	\begin{center}
 		\resizebox{0.5\textwidth}{!}{%
 			\includegraphics{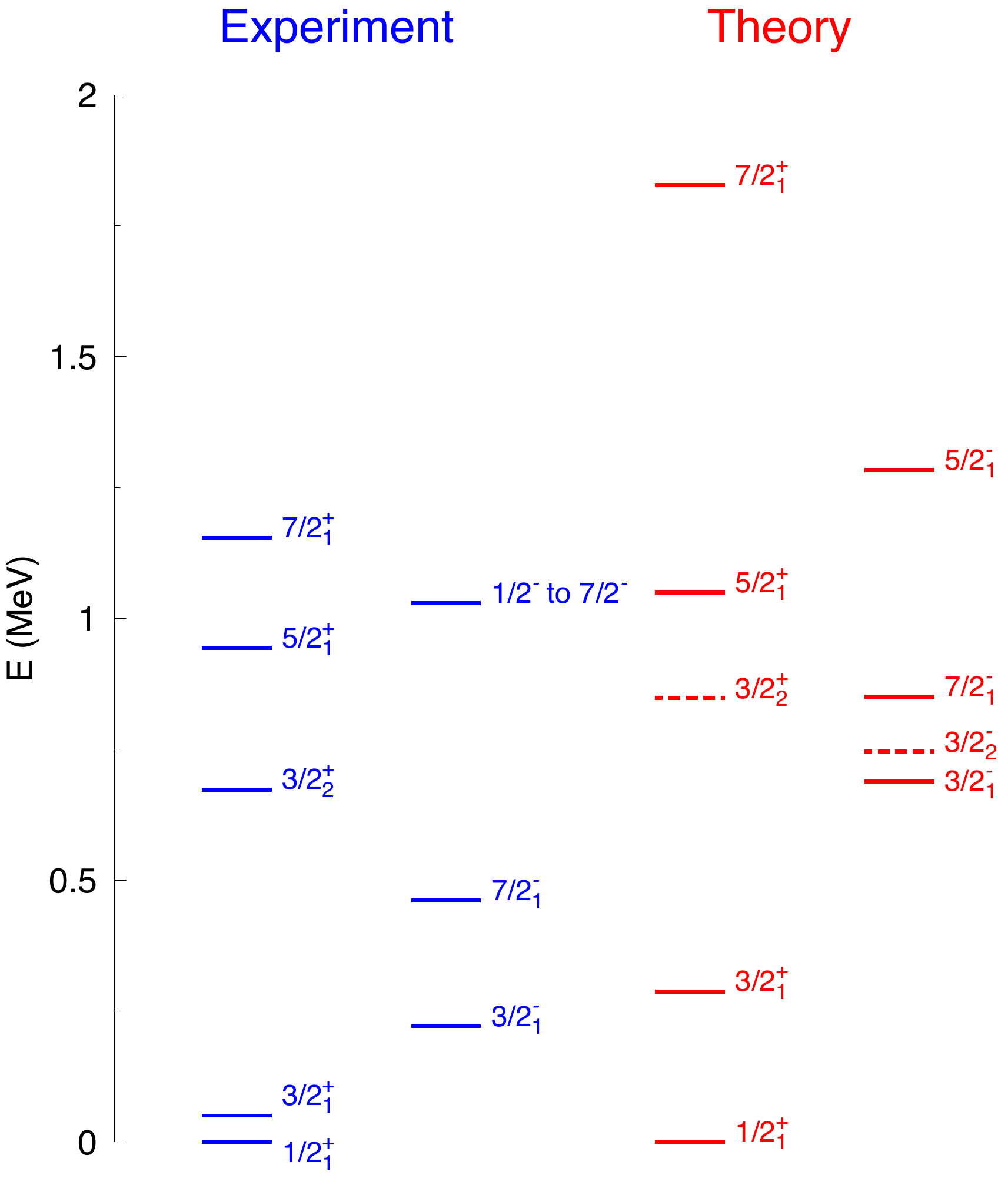}}
 		\caption{Spectrum of the nucleus $^{31}$Mg in the symmetry conserving mean field approximation.}
 		\label{Fig:spectrum}       
 	\end{center}
 \end{figure}
 
\section{Spectrum and other observables}
\label{Sect:spec_plus}
 The solution of the Hill-Wheeler-Griffin equation, Eq.~(\ref{HW_K}), as a function of $(\beta,\gamma)$ provided the PESs displayed in Fig.~\ref{Fig:PES_PNAMP} for  the different values of the angular momentum.  Though in each $(\beta,\gamma)$ point there are $(2I+1)/2$ linearly independent solutions, the energies displayed in the PESs  correspond to the the lowest solution at each point.  The minima of these surfaces provide the energies of the states  corresponding to the spin and parity of the given PES. Since, within a PES, the wave functions of the different  $(\beta,\gamma)$ points, in general, are not orthogonal to each other, only one point provides a physical state.
In Fig.~\ref{Fig:spectrum} we present the excitation energies predicted by the minima of these surfaces. The excited states obtained by considering the solutions $\sigma=2$ or higher at each point $(\beta,\gamma)$ are not considered in this work.  The only exception are the states ${3}/{2}^{+}_2$  and ${3}/{2}^{-}_2$ represented by dashed lines in  Fig.~\ref{Fig:spectrum} corresponding to the secondary minimum,  at $(\beta=0.31, \gamma=14.^{\circ})$  and $(\beta=0.69, \gamma=12.^{\circ})$ of Fig.~\ref{Fig:PES_PNAMP}, respectively.   We have exceptionally included them for two reasons: First because these are the levels that should be lowest  in the case that inversion has not taken place and second because they are practically orthogonal to the other  $I={3}/{2}_{1}$  states present in the corresponding PES.

In Fig.~\ref{Fig:spectrum} we have also plotted the experimental spectrum \cite{ENSDF,Ne.11}.  Concerning the positive parity part of the spectrum the ordering of the levels is correctly reproduced by our calculations but rather stretched. As mentioned in the Introduction the approach we are presenting here is the symmetry conserving mean field approach and it is well known  from even-even nuclei \cite{Nuria,Nuria2,BRE.15,EBR.16} that residual correlations will compress the spectrum. With respect to the negative parity part of the spectrum, experimentally there are two levels clearly identified, the ${7}/{2}^{-}_1$ and the ${3}/{2}^{-}_1$. In the theory they appear again in the right order but too high in energy. One expects again that the residual interactions will compress the spectrum.

Concerning the electromagnetic moments they are quoted in Table~\ref{table2} for the positive and negative bands.   Our result for the  magnetic moment of the ground state is relatively close to the measured value   $-0.88355(15) \mu_N$  \cite{moma}. In the calculations we have used the free gyromagnetic values. The electric quadrupole moments correspond to a well deformed nucleus, unfortunately there are no experimental values for this observable.  In Table~\ref{table3} we have listed experimental values for dipole and quadrupole electromagnetic transitions and a selection of the theoretical predictions. With the exception of the $M1$ transition from the $\frac{3}{2}^{+}_{1}$ level to the $\frac{1}{2}^{+}_{1}$ we obtain in general a very reasonable agreement with the experimental values.
   
   \begin{table}
   	\centering
   	\begin{tabular}{|c|c |c| c| c|c|}
   		
   		\hline
   		$\footnotesize I^{\pi}_{\sigma}  $  &  $Q_{spec}$ &  $\mu$ &   $ I^{\pi}_{\sigma}$ &   $Q_{spec}$&  $ \mu$
   		\tabularnewline
   		\hline
   		${1}/{2}^{+}_{1}$  &  $0$ &   $-0.935$ &        ${1}/{2}^{-}_{1}$ &  $0$  & $1.339$
   		\tabularnewline
   		${3}/{2}^{+}_{1}$  &  $-15.5$ &   $0.690$ & ${3}/{2}^{-}_{1}$ & $-12.16$  & $-1.594$
   		\tabularnewline
   		${3}/{2}^{+}_{2}$  &  $8.40$ &   $1.132$ &       ${3}/{2}^{-}_{2}$  & $16.63$  & $-0.695$
   		\tabularnewline
   		${5}/{2}^{+}_{1}$  &  $-20.00$ &   $-0.085$ &${5}/{2}^{-}_{1}$ &   $-6.17$ & $-0.125$
   		\tabularnewline
   		${7}/{2}^{+}_{1}$  &  $-26.34$ &   $1.557$ &  ${7}/{2}^{-}_{1}$ &   $-21.50$ & $-0.910$
   		\tabularnewline
   		\hline
   	\end{tabular}
   	\caption{Electromagnetic moments of the positive and the negative parity bands.  $Q_{spec}$  is given in units of ${\rm efm}^{2}$and  $\mu$  is in units of $\mu_{N}$.}
   	\label{table2}
   \end{table}

\begin{table}
	\centering
	\begin{tabular}{|c|c |c|}
		
		\hline
		$\footnotesize I^{\pi}_{\sigma}  $  &  Experiment &  Theory
		\tabularnewline
		\hline
		$B(E2; 5/2^{+}_1\longrightarrow 1/2^{+}_1)$  &  61(7) \cite{exp1}&   117
		\tabularnewline
		\hline
		$B(M1; 5/2^{+}_1\longrightarrow 3/2^{+}_1)$  &  0.1-0.5 \cite{exp1}&   0.590
		\tabularnewline
		\hline
		$B(M1; 3/2^{+}_1\longrightarrow 1/2^{+}_1)$  &  0.019(0.004) \cite{exp2}&  0.093
		\tabularnewline
		\hline
		$B(E2; 7/2^{-}_1\longrightarrow 3/2^{-}_1)$ &  68(5) \cite{exp3}&  88 
		\tabularnewline
		\hline
		$B(E2; 3/2^{+}_1\longrightarrow 1/2^{+}_1)$ &  $-$ &   109
		\tabularnewline
		\hline
		$B(E2; 5/2^{+}_1\longrightarrow 3/2^{+}_1)$ &  $-$ &   38
		\tabularnewline
		\hline
		$B(E2; 7/2^{+}_1\longrightarrow 5/2^{+}_1)$ &  $-$ &   13
		\tabularnewline
		\hline
		$B(E2; 7/2^{+}_1\longrightarrow 3/2^{+}_1)$ &  $-$ &   146
		\tabularnewline
		\hline
		$B(M1; 7/2^{+}_1\longrightarrow 5/2^{+}_1)$ &  $-$ &   0.107
		\tabularnewline                
		\hline
	\end{tabular}
	\caption{Transition probabilities in $^{31}$Mg, $B(M1)$  in units of $\mu_{N}^{2}$  and $B(E2)$  in ${\rm e}^{2}{\rm fm}^{4}$}
	\label{table3}
\end{table}

\section{Conclusions}
\label{Sect:concl}

We have presented a  calculation for odd-nuclei in the  framework of the symmetry conserving mean field approach. 
We have applied it to the description of the nucleus $^{31}$Mg at the border of the inversion island. We find the two coexisting minima corresponding to the normal occupation and the intruder one typical for the inversion island.

In spite of the simplicity of the approximation we obtain a qualitative  agreement of the theoretical and the experimental spectrum. As expected due to the lack of correlations we obtain a stretched spectrum.  The experimental values of the measured  electromagnetic transitions as well as the magnetic moment are also well described.

The results obtained for the nucleus  $^{31}$Mg with the Gogny force encourage us  to improve the present approach to achieve the same degree of sophistication as for even-even nuclei. 

\section*{Acknowledgements}
This work was supported by the Ministerio de Econom\'ia y Competitividad under contracts No. FPA2011-29854- C04-04, No. FPA2014-57196-C5-2-P.

%

\end{document}